# Resting state fMRI-based temporal coherence mapping

Abbreviated title: temporal coherence mapping

Ze Wang, PhD

ORCID: 0000-0002-8339-5567

Department of Diagnostic Radiology and Nuclear Medicine

University of Maryland School of Medicine

670 W. Baltimore St, Baltimore, MD 20201

ze.wang@som.umaryland.edu

Conflict of interest: none

Data availability statement: the HCP data is freely available from the HCP consortium.

Ethics approval statement: data reanalysis has been approved by IRB. Patient consent forms were obtained by HCP.

**Acknowledgements**

The research effort involved in this study was supported by NIH grants: R01 AG070227, R01AG081693, R01AG060054, AG060054-02S1, R01EB031080-01A1, R21AG08243, and by the support of the University of Maryland, Baltimore, Institute for Clinical & Translational Research (ICTR) and the National Center for Advancing Translational Sciences (NCATS) Clinical Translational Science Award (CTSA) grant number 1UL1TR003098. Both imaging and behavior data were provided by the Human Connectome Project, WU-Minn Consortium (Principal Investigators: David Van Essen and Kamil Ugurbil; 1U54MH091657) funded by the 16 NIH Institutes and Centers that support the NIH Blueprint for Neuroscience Research; and by the McDonnell Center for Systems Neuroscience at Washington University in St. Louis. The author thanks the Human Connectome Project for open access to its data, thanks Donghui Song, Qiu Ge, Yiran Li for data organization, transferring, and processing assistance.

Abstract

Long-range temporal coherence (LRTC) is a common feature of a self-organized dynamic system, which is fundamental to the system function. LRTC in the brain has been shown to be important to cognition. Assessing LRTC may provide critical information for understanding the potential underpinnings of brain organization, function, and cognition. To facilitate this overarching goal, I provide a method, which is named temporal coherence mapping (TCM), to explicitly quantify LRTC using resting state fMRI. TCM is based on correlation analysis of the transit states of the phase space reconstructed by temporal embedding. A few TCM properties were collected to measure LRTC, including the averaged correlation, anti-correlation, the ratio of correlation and anticorrelation, the mean coherent and incoherent duration, and the ratio between the coherent and incoherent time. TCM was first evaluated with simulations and then with the large Human Connectome Project data. Evaluation results showed that TCM metrics can successfully differentiate signals with different temporal coherence regardless of the parameters used to reconstruct the phase space. In human brain, TCM metrics except the ratio of the coherent/incoherent time showed high test-retest reproducibility; TCM metrics are related to age, sex, and total cognitive scores.  In summary, TCM provides a first-of-its-kind tool to assess LRTC and the imbalance between coherence and incoherence; TCM properties are physiologically and cognitively meaningful.

## Introduction

The human brain is a self-organized system(Haken, 2012; Singer, 2009; Willshaw, 2006) which relies on the time varying activity to coordinate and adapt its functions. Understanding the temporal dynamics of brain activity is crucial to understanding the individual differences of brain function and the neuropsycho-pathologies associated with neuropsychiatric conditions. A critical aspect of these temporal dynamics is the emergence of long-rang temporal coherence or correlation (LRTC), meaning that brain activity at one moment can influence future activities. LRTCs have been well observed at different time scales using neuronal spike recordings, electrophysiological data, hemodynamic response as measured by functional MRI(He, 2011; He et al., 2010; Suckling et al., 2008; Wink et al., 2008). They have been shown to be crucial to high-order brain functions including decision making, memory, learning, network reorganization (Botcharova et al., 2014), attention, perception, coordination, etc (Buschman and Miller, 2007; Buzsáki and Draguhn, 2004; Dean et al., 2012; Palva et al., 2013; Pesaran et al., 2002; Saleh et al., 2010; Shewcraft et al., 2020; Thut et al., 2012; Womelsdorf et al., 2006a, b; Wong et al., 2016). LRTC of a single brain region may initiate or influence cross-regional effective connectivity and communications (Fries, 2005, 2015; Lu et al., 2017; Pesaran et al., 2018; Teki et al., 2013).

Given the importance of LRTC in functional brain organization and neurocognition, several methods have been used to quantify it. The most intuitive one is the auto-correlation function (ACF) since a slowly decaying ACF indicates LRTC. In the case where ACF is not reliable to estimate, LRTC can be assessed with the Hurst exponent (Hurst, 1951) or its surrogates such as the exponent of the apparent power law function fitted from the logarithm of the spectrum or the linear line slop estimated by the detrended fluctuation analysis (DFA) (Peng et al., 1995). While these methods have been proven to be useful for evaluating LRTCs, none of them directly considers the transit states of the underlying dynamic process. According to theoretical work and neuroscience experiments, the human brain nearly operates at a self-organized criticality, from where the LRTCs emerge (Deco and Jirsa, 2012; Rubinov et al., 2011). At the critical condition, the dynamic system frequently switches between different intermediate state around the attracting point(Beggs and Plenz, 2003; Friedman et al., 2012; Munoz, 2018) and LRTCs emerge from the inter-state correlations across a long time range and can be accordingly directly characterized through the inter-transit state correlations. During the course, I can separate correlated states from anti-correlated transit states, so that I can directly evaluate the macroscopic level brain temporal coherence and anti-coherence (C/A) balance. Considering the fact of that at the microscopic level, it is well known that neurons preserves a neuronal excitation and inhibition (E/I) balance that is important to neuron function (Okun and Lampl, 2008), the macroscopic level C/A balance may carry important information about brain self-regulation and brain function.

Phase space is mathematically defined as a collection of all states of a system. It is widely used to study the behavior of the time evolution trajectory of the system status and to use current state to predict future behavior. According to Takens' theorem (Takens, 1981), the phase space can be reliably reconstructed from the measure one-dimensional timeseries through temporal embedding. Each embedding vector identified during phase space reconstruction was considered representing an intermediate state. We have previously adopted the temporal embedding-based approximate entropy as a tool to study coherence of resting state fMRI (Wang et al., 2014). The approximate entropy, the so-called Sample Entropy (SampEn) (Richman and Moorman, 2000) is calculated by the negative logarithm of the probability of that two similar phase space transit states(Nolte, 2010) will remain similar if the dimension of reconstructed phase space increased by one, meaning that the length of the embedding vector increases by one. While the SampEn-based entropy mapping has been shown to be informative of brain health and neurocognitions and sensitive to neuromodulation, medication, or nonpharmacological modulation (Brian N Lee, 2023; Camargo et al., 2024; Da Chang, 2018; Del Mauro et al., 2024; Del Mauro and Wang, 2023, 2024a, b; Donghui Song, 2019; Jiang et al., 2023; Jordan et al., 2024; Li et al., 2016; Liu et al.,

2024; Liu et al., 2020; Song et al., 2024; Wang, 2009, 2020; Wang, 2021b; Wang and Initiative, 2020; Xue et al., 2019; Ze Wang, 2017; Zhao et al., 2024; Zhou et al., 2016), it only provides an indirect way to assess temporal coherence. Moreover, it does not consider temporal anti-coherence, not to mention the C/A balance. To bridge this gap, I proposed a new method: temporal coherence mapping (TCM), to directly quantify LRTC of a dynamic system through the correlations between all intermediate states of the phase space of a dynamic system (Nolte, 2010). The inter-state correlation was calculated and averaged across all possible pairs of embedding vectors. By aggregating the positive and negative correlations separately, I also provided a way to examine the temporal C/A ratio. Although TCM is still based on temporal embedding as in our previous entropy mapping work but it differs from the former by directly characterize the two-dimensional relationship between the embedding vectors (each dimension is the time step of the embedding vector). Temporal embedding was based on a small window of 2 or 3 timepoints in our previous brain entropy mapping work. By contrast, I used much longer embedding vectors in TCM so that inter-vector correlations can be reliably estimated. Temporal coherence was then measured as the mean positive correlation coefficient of those embedding vectors and temporal anticoherence was by the mean negative correlation coefficient. Their ratio was used to assess the C/A balance. Another C/A balance measure was defined by the ratio between the average length of the diagonal line segments of the positive correlation coefficient matrix and that of the diagonal line segments of the negative correlation coefficient matrix. The diagonal lines in the correlation coefficient matrix of the embedding vectors indicate how long the dynamic system stays in certain transit states and are useful for indicating periodic behavior of the system(Marwan et al., 2007).

The two-dimensional similarity matrix of the embedding vectors-derived phase space can be characterized by the recurrence plot-based quantitative analysis (RQA)(J.-P. Eckmann, 1987), which has been applied to task fMRI analysis in the literature (Bianciardi et al., 2007; Lombardi et al., 2019; Lombardi et al., 2015; Lopes et al., 2021; Rangaprakash et al., 2013). In RQA, the embedding vector is often limited to a few consecutive timepoints and similarity between embedding vectors is calculated with a distance metric and then thresholded to be a binary number based on an arbitrary cutoff. Our TC analysis differs from RQA by using a longer time window to extract the transit status of the phase space and using the correlation coefficient to measure the similarity and by characterizing both the non-thresholded and thresholded similarity matrix, which carry the following benefits: the use of long window allows a more reliable estimation of the similarity; the use of correlation coefficient as the similarity measure avoids the influence of the signal intensity which may drift along time; the non-thresholded matrix property is independent of the cutoff; using correlation coefficient as the similarity metric allows investigating both the coherence and anti-coherence and their balance at the same time.

I applied TCM to brain activity measured with resting state functional MRI (rsfMRI) and empirically examined the temporal brain C/A balance based on these TCM C/A metrics at the macroscopic level for the first time. My hypothesis was that resting healthy brain presents spatially distributed LRTCs with higher LRTCs in grey matter than in white matter; the long-range positive correlations and long-range negative correlations are well balanced in the brain as reflected by a C/A ratio <1 across the brain. I also hypothesized that LRTCs present regional correlations to physiological measures and neurobehavior measures.

## Materials and Methods

### *Ethics statement*

Data acquisition and sharing have been approved by the HCP parent IRB. Written informed consent forms have been obtained from all subjects before any experiments. This study re-analyzed the HCP data and data Use Terms have been signed and approved by the WU-Minn HCP Consortium.

*Data included*

rsfMRI data, demographic data, and neurobehavior data from 1102 healthy young subjects were downloaded from HCP. After excluding subjects who did not have full rsfMRI scans, or demographic, or behavioral, or physiological data, 865 remained (age 22-37 yrs, male/female=401/464). The range of education years was 11-17 yrs with a mean and standard deviation of 14.86±1.82 yrs. The rsfMRI data used in this paper were the extended processed version released on July 21 2017. Each subject had four rsfMRI scans acquired with the same multi-band sequence(Moeller et al., 2010) but the readout directions differed: readout was from left to right (LR) for the 1st and 3rd scans and right to left (RL) for the other two scans. The purpose of acquiring different scans with opposite phase encoding directions was to compensate the long scan time induced image distortion. MR scanners all present field strength (B0) inhomogeneity, which causes signal distortion because of the imperfect excitation using the radiofrequency pulses that are tuned to the frequency determined by the ideal B0. While the B0 inhomogeneity caused distortions can be well corrected using two additionally acquired calibration scans using the opposite phase encoding directions: one is with LR and the other is with RL, HCP acquired two LR and two RL rsfMRI scans for the purpose of assessing the potential residual effects after the distortion correction and to assess the test-retest stability of rsfMRI measure. Each scan had 1200 timepoints. Other acquisition parameters for rsfMRI were: repetition time (TR)=720 ms, echo time (TE)=33.1ms, resolution 2x2x2 $mm^3$. The pre-processed rsfMRI data in the Montreal Neurological Institute (MNI) brain atlas space were downloaded from HCP (the S1200 release) and were smoothed with a Gaussian filter with full-width-at-half-maximum = 6mm to suppress the residual inter-subject brain structural difference after brain normalization and artifacts in rsfMRI data introduced by brain normalization. Non-neural spatiotemporal signal components were removed using the ICA-FIX algorithm(Griffanti et al., 2014; Salimi-Khorshidi et al., 2014; Smith et al., 2013). Motion parameters and their derivatives were regressed out from the time series too. More preprocessing details can be found in the HCP data release manual.

*Temporal coherence mapping (TCM)*

Phase space reconstruction through temporal embedding is illustrated in the top panel in Fig. 1. Denote a time series, for example, the time series of a brain voxel, by $\mathrm{x} = [\mathrm{x}_1, \mathrm{x}_2, \ldots \mathrm{x}_\mathrm{N}]$, where N is the number of time points. The phase space of the underlying dynamic system can be reconstructed by a series of embedding vectors, each with w consecutive points extracted from x: $\mathbf{u}_\mathrm{i} = [\mathrm{x}_\mathrm{i}, \mathrm{x}_{\mathrm{i}+1}, \ldots \mathrm{x}_{\mathrm{i}+\mathrm{w}-1}]$, where $\mathrm{i} = 1$ to N-w+1, w is the pre-defined embedding vector length. As illustrated by the lower panel of Fig. 1, for a specific time series x, TCM is to calculate the correlation coefficient matrix of the embedding vectors. For the simplicity of description, this matrix was named the TCM matrix in the following text.

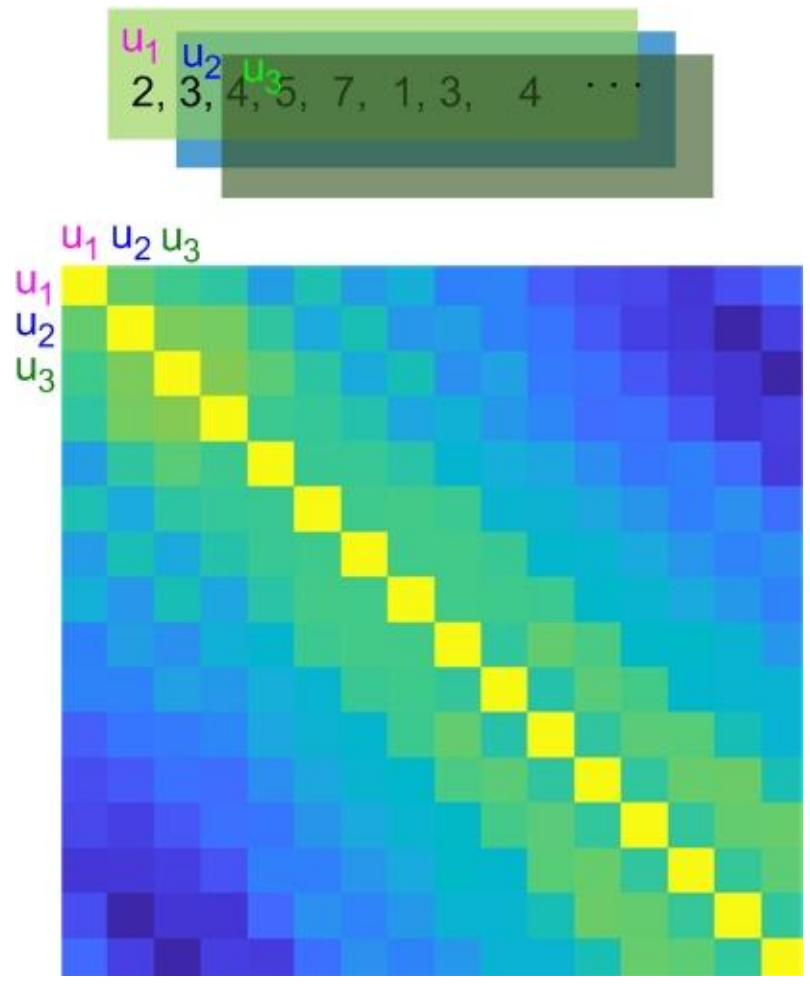

Figure 1. Illustration of TCM. The upper panel shows the moving window-based embedding vector extractions; the lower panel shows the TCM matrix (the correlation coefficient matrix) of those embedding vectors. Yellow means a correlation coefficient of 1; blue means negative correlations. $u_1$, $u_2$, $u_3$ denote three embedding vectors.

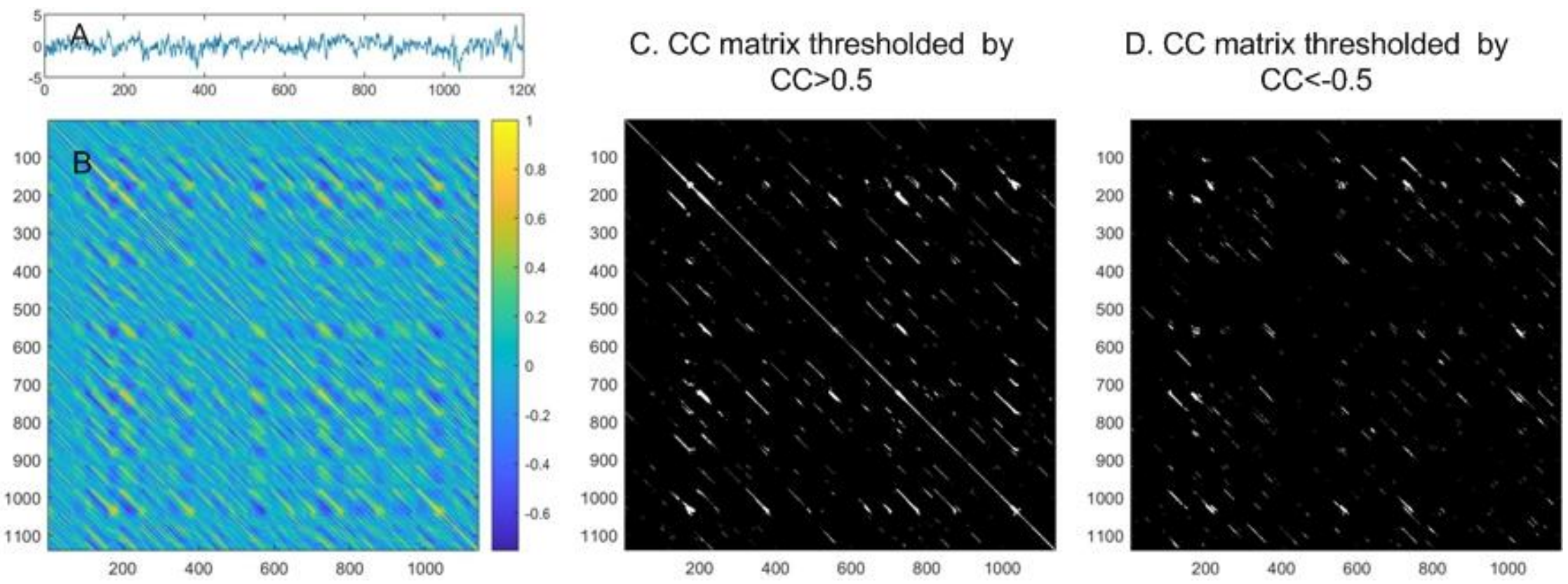

Fig. 2. The TCM matrix of the embedding vectors of a rsfMRI time series extracted from a representative HCP rsfMRI scan in the precuneus. A) The rsfMRI time series, B) The non-thresholded correlation coefficient matrix, C and D are the binarized TCM matrix thresholded by CC>0.5 and CC<-0.5, respectively.

While the TCM matrix provides a way to observe the temporal coherence patterns such as the positive and negative correlation, the decaying and potentially recurring correlation from the main diagonal to the off-center diagonals, the balance between the positive and negative correlations etc, I need some metrics to quantify these properties. Below, I provided several potentially valuable property metrics for condensing the information provided by the TCM matrix.

(1) The first two are the temporal coherence (TC) and the temporal anti-coherence (TAC) which are calculated as the mean positive correlation coefficient and the mean negative correlation coefficient:

$$TC = \frac{\sum_i^M cc_i H(cc_i)}{M} \quad \text{for all } cc_i > 0$$

$$TAC = \frac{\sum_i^M -cc_i H(-cc_i)}{M} \quad \text{for all } cc_i < 0$$

where M=(N-w+1)*(N-w) is the total number of off-diagonal elements of the correlation coefficient matrix. H(.) is the Heaviside function:

$$H(x) = \begin{cases} 1, x > 0 \\ 0, \ \ x \le 0 \end{cases}$$

(2) The third is the *C/A balance measure 1 (CAB1)*: TC-TAC.

(3) The last three measures are based on the mean length of the continuous diagonal line segments of the binarized TCM matrix. The mean length of the continuous diagonal lines is defined in the recurrent plot analysis for measuring the divergency behavior of the dynamic system (Marwan et al., 2007). In this paper, I calculate this length for the positive TCM matrix and negative TCM matrix separately. As shown in Fig. 2, the TCM matrix can be binarized with a threshold to only retain the positions with stronger than the correlation strength cutoff. Isolated dots in the binarized matrix indicate rare transit states. In other words, they do not persist in time, or they simply fluctuate too much. By contrast, the continuous diagonal line segments suggest that the system revisits the transit state represented by the embedding vectors at the corresponding coordinates of these segments many times. The length of those diagonal line segments provides an estimate of how long these transit states stick together. By hard-thresholding the TCM matrix with a positive threshold r, I can get a binarized positive correlation coefficient matrix (1 means CC>r, 0 means CC<=r). A negative threshold -r can be used to get a binarized negative TCM matrix (1 means CC<-r, 0 means CC>=-r). The following two measures can be used to measure the recurrence of coherent (positively correlated) and incoherent (negatively correlated) transit states, separately. Their ratio can be taken as another C/A balance indicator.

*MLP*: the mean length of the diagonal line segments of the binarized positive TCM matrix.

*MLN*: the mean length of the diagonal line segments of the binarized negative TCM matrix.

*MLP-MLN*

**Algorithm 1. TCM**

```
    input:
      x - time series with N timepoints
      w - embedding vector length
      r - correlation coefficient threshold
      g - gap between adjacent embedding vectors
    output:
      TC: mean positive correlation coefficient
      TAC: mean negative correlation coefficient
      CAB1: TC-TAC
      MLP: mean length of the diagonal lines of
          positive TCM matrix
      MLN: mean length of the diagonal lines of
          negative TCM matrix
      CAB2: MLP-MLN
    Definitions of intermediate variables: cc: correlation coefficient,
    DIA_S/E: the starting and end diagonals used for MLP and MLN
    calculations.  seg_e/seg_s: the starting and end position of a
    continuous diagonal segment. pseglen/nseglen: length of the
    diagonal segments for the positive and negative correlations
    # line 1 to line 8 are to initialize several variables.
 1  DIA_S=w/3
 2  DIA_E=w
 3  Nv=int((N-w+1)/g)  #number of embedding vectors
 4  max_dia=(Nv-DIA_E)*g   #maximal diagonal number
 5  M=N-w-1
 6  Totnum_cc = Nv*(Nv-1)/2-DIA_E*(DIA_E+1)/2
 7  TC=TAC=0
 8  MLP=MLN=0
 9
10  for (l=DIA_S*g; l<max_dia; l+=gap)
11    tlen_end=tlen-w-l
12    #reset flags for detecting a diagonal segment
13    seg_new=0;            #flag: 0 means found a new segment
```

```
        seg_s=0;    seg_e=0;  #the start and end position in the
      #diagonal of current diagonal segment
14      for(i=0; i < tlen_end; i=i+gap)
15          find embedding vectors from x at position i and i+l*gap
16          calculate their correlation cc
17          if(cc>0) add cc to TC
18          else     add cc to TAC
19          if(cc>r)
20              if(seg_new==0)  # find a new diagonal segment
21                # record the start point and set the flag
22                seg_s=i;   seg_e=i+1; seg_new=1;
23              else          # continuation of an existing line
24                seg_e+1 -> seg_e
25          else
26              # set the end of the segment if needed
27              if(seg_new==1)
28                  seg_new=0; pseglen=seg_e-seg_s;
29                  # exclude the isolated point
30                  if(pseglen>1)
31                      MLP = MLP+pseglen;
32          if(cc<-thr)
                if(nseg_new==0)  # find a new diagonal segment for the
33    negative correlation
                    nseg_s=i;   nseg_e=i+1; nseg_new=1;  #nseg_s and
34    nseg_e are the starting and end position of the current segment
35              else
36                  nseg_e+1 -> nseg_e;
37          else
38              if(nseg_new==1)
39                  nseg_new=0; nseglen=nseg_e-nseg_s;
40                  if(nseglen>1)
41                       MLN = MLN+nseglen;
42      // force the last diagonal line to stop
43      if(seg_new==1)
44          seg_new=0;
45          pseglen=seg_e-seg_s;
46          if(pseglen>1)
47               MLP = MLP+pseglen;
48      if(nseg_new==1)
49          nseg_new=0;
50          nseglen=nseg_e-nseg_s;
51          if(nseglen>1)
52              MLN = MLN+ nseglen;
53    // end of the l loop
54    MLP =MLP/ Totnum_cc;
```

```
55  MLN =MLN/ Totnum_cc;
56  CAB2=MLP-MLN;
57  TC =TC/ Totnum_cc;
58  TAC=TAC/ Totnum_cc;
59  CAB1=TC-TAC;
```

*Algorithm and implementation*

For a single voxel with a length of T timepoints, the computation complexity of TC/TAC/CAB1 is approximately $O((T-w+1)+(T-w)^2 \cdot w^2)$. For N voxels, the computation complexity will be multiplied by N. To accelerate the process, I used parallel computing. Meanwhile, calculating the entire TCM matrix and then the parameters requires large computer memory. To avoid this burden, I calculated the TCM properties on the fly. Because the TCM matrix is symmetric, I only considered the upper triangle matrix to save time. The main computation loop is for different delay. In the TCM matrix, each diagonal corresponds to a specific delay between the assessed two embedding vectors. By looping over the diagonals, I managed to calculate the aforementioned TCM matrix properties without pre-calculating the entire matrix. Detailed algorithm for TCM and the property calculation was given in Algorithm 1. "gap" is an integer here for increasing the time interval between adjacent embedding vectors. In fMRI, embedding vectors with a few timepoints away often show very high correlation coefficients due to the hemodynamic response function convolution. As a result, the binarized TCM matrix has many long lines in the diagonals within the nearest neighborhood of the main diagonal. These diagonals may even appear as entire continuous lines and will dominate the calculation of the mean length of the diagonal lines and result in a nearly constant value across different subjects. To avoid their influence, I introduced a parameter DIA_S to exclude the diagonals near the main diagonal from the several TCM parameter calculations. This number has been empirically verified to be between 1/4w and 1/2w. DIA_E is used to exclude the last several diagonals because they are relatively too short to provide sufficient points to find reliable continuous line segments.

Two types of parallel computing were used. One was based on CUDA (the parallel computing programming platform created by Nvidia Inc). Similar to the brain entropy mapping (Del Mauro and Wang, 2024a; Wang, 2021b; Wang et al., 2014) CUDA acceleration, I used parallelism across within-brain voxels. The second was multiple threads. The former was much faster than the latter but was limited by the availability of graphic process unit cards. For the HCP rsfMRI data, I used the multi-thread version as it could be used in our massive high performance computing system.

*Experiments*

In general, the embedding vector length (w) should be kept short in order to capture more transit information. Big w will reduce the overall coherence because correlation coefficient of two different vectors tends to decrease with the vector length. However, the reductions of coherence and anti-coherence due to the change of w may not be linearly related, which will lead to variability of the coherence to anti-coherence ratio. To evaluate this effect, I generated sinusoidal signals with different frequencies and used different w to calculate the ratio between coherence and anti-coherence through CAB1 and CAB2. To calculate CAB2, I used a threshold of $|r|>=0.3$ to binarize the temporal coherence matrix and calculate CAB2 using the algorithm described above. I used sinusoids because a single sinusoidal is periodic and fully balanced. In other words, I know that the gold standard of CAB1 and CAB2 of any sinusoidal should be 1.

To evaluate the effects of w and r on temporal coherence characterization for difference signals, I extracted mean time series from the first rsfMRI scan (the first LR scan) of 20 HCP subjects from the posterior cingulate cortex (PCC). 20 1/f noise and gaussian noise were also generated. TCM was performed using the above algorithm with different w and r values: w varied from 30 to 90 with a step of 10; r varied from 0.2 to 0.6 with a step of 0.1. TC, TAC, CAB1, MLP, MLN, and CAB2 were calculated. Analysis of variance (ANOVA) and paired-t test were used to statistically infer the effects of w when r was fixed or the effects of r when w was fixed.

I then calculated TC, TAC, CAB1, MLP, MLN, and CAB2 at each voxel for the 862 HCP subjects four rsfMRI data. Four Nvidia GTX 1080 Titan graphic processing unit (GPU) video cards were used to accelerate the process. The collections of each measure at all voxels form a corresponding map, which was called a temporal coherence map (TCM). For each subject, each of the six TCMs was averaged across the first LR and the first RL rsfMRI scans to minimize the potential effects of the phase encoding polarities. For the simplicity of description, the mean TCMs were called REST1 TCMs. The same averaging process was performed for these parametric maps calculated from the second LR and the second RL rsfMRI scans, and I called the mean TCMs as REST2 TCMs. w=30, 60, 90, and r=0.3 and 0.5 were used.

*Test-retest stability*

TCM test-retest stability was assessed by the intra-class correlation (ICC) (Shrout and Fleiss, 1979) between the corresponding REST1 and REST2 TCMs, i.e., for each of the six TCMs separately.

*Statistical analyses on the biological and cognitive associations of TCMs.*

To find the potential biological or neuropsychological associations of resting brain TCM properties, I performed several voxelwise regression analyses for each of the six TCMs collected at the REST1 session (averaged across the first LR and the RL scans) and the REST2 session (averaged across the second LR and the RL scans), separately. Biological measures included age and sex. Cognitive capability was measured by the total cognitive function composite score (CogTotalComp_Unadj) in the NIH toolbox (http://www.nihtoolbox.org) that is derived by averaging the normalized scores of each of the included fluid and crystalized cognition measures and then deriving scale scores based on the new data distributions. Higher scores mean higher levels of cognitive functioning. I used the fully processed rsfMRI data provided by the HCP consortium and I have recently demonstrated that the full processing including the noise component removal step successfully suppressed nuisance effects related to physiological confounds such as respirational fluctuations and cardiac cycles(Del Mauro and Wang, 2023). To further control the residual respiratory and cardiac effects, I included the respiration rate (RR) and heart rate (HR) as two additional nuisance variables in all regression models. RR and HR were calculated from the corresponding record for each rsfMRI scan session. Signal sampling rate of those records was 400 Hz, which is roughly 288 times faster than the sampling rate of rsfMRI. Both time series were low pass filtered with a cutoff of 5 Hz and 10 Hz for the respiration and cardiac data, respectively. Local maxima were then detected using Matlab (Mathworks, Natick, Massachusetts, United States) function islocalmax (Matlab 2021b). The first derivative of the time stamps of the local maxima was calculated. The rmoutliers function of Matlab was used to remove outlier peak to peak time differences. The remaining time differences of each recording were averaged and considered to be the respiration cycle and heart beat cycle, respectively. Subjects were excluded from the following analyses if their final mean respiration cycle was longer than 10 secs or if the heart beat cycle was longer than 4 secs. RR and HR of the two scans of each scan session were averaged and paired with the corresponding TCM maps of each scan session (the mean of the LR and RL scans).

Two regression models were built. The first included sex, age, RR, and HR in the model and was used to investigate TCM vs sex and age correlations. The second included sex, age, RR, HR, education years, and total cognitive score as covariate and was used to study the correlation between TCM and the total cognition. Because total cognition score was collected in the same date as the second rsfMRI session, mean LR and RL TCMs of REST2 were used in the second regression model. The multiple regression model was built and estimated using Nilearn (https://nilearn.github.io/).

The voxelwise significance threshold for assessing each of the association analysis results was defined by $p<0.05$. Multiple comparison (across voxels) correction was performed with the family wise error theory (Nichols and Hayasaka, 2003) or false detection rate ($q<0.05$). Image and statistical results were displayed using Mricron (https://www.nitrc.org/projects/mricron) developed by Chris Rorden.

## Results

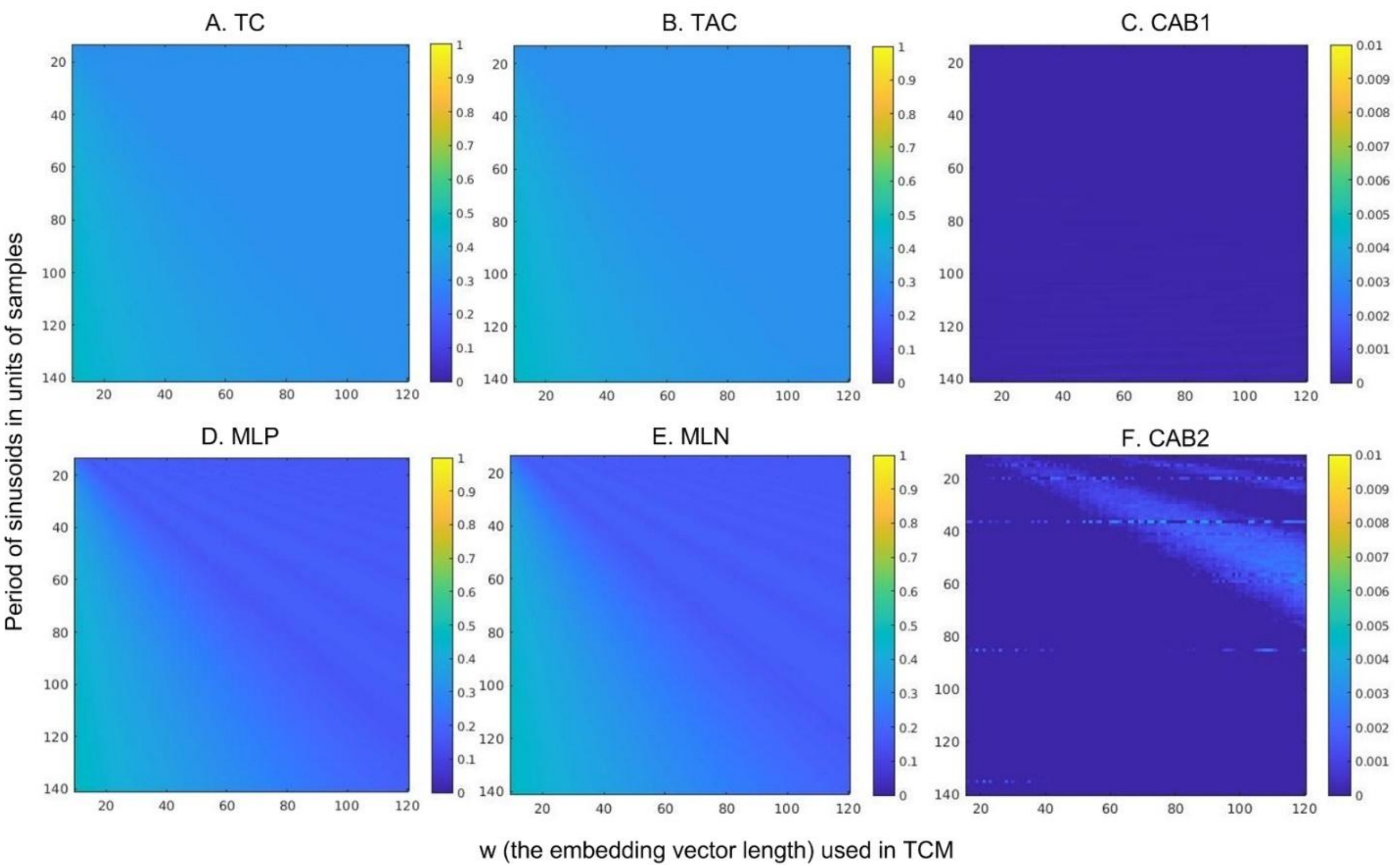

Fig. 3. TC parameters of sinusoids with different frequencies. The y axis of each subfigure is the period of the sinusoid. The x axis is the embedding vector length w. r=0.3 was used to calculate

MLP, MLN, and CAB in D, E, F, respectively. Colorbar on the right side of each subfigure depicts the display window for the corresponding TC map shown on the left of each subfigure.

*Interacting effects of w and signal frequency on TC parameters*

Fig. 3 shows the dependency of TC parameters on the sinusoid frequencies (period was shown in Fig 3). At the same period (in the unit of samples), TC (mean temporal correlation) and TAC (mean temporal anticorrelation (negative correlation)) of the sinusoid decrease with the embedding vector length and stop decreasing after w is longer or equal to the period. At each embedding vector length, TC, TAC, MLP, MLN increase with the period of the sinusoid (decrease with the frequency of the sinusoid). The frequency and embedding vector length effects were mostly suppressed in CAB1 and CAB2. TC and TAC showed less variations compared to MLP and MLN and the variations of MLP and MLN were moderate (~2%). CAB1 is nearly 0 in the entire assessed frequency and w values; CAB2 showed small pseudo-periodic frequency vs w interaction effects when w is longer than signal cycles. CAB2 also showed some w independent fluctuations in a few frequencies (the horizontal discontinous lines in Fig. 3F).

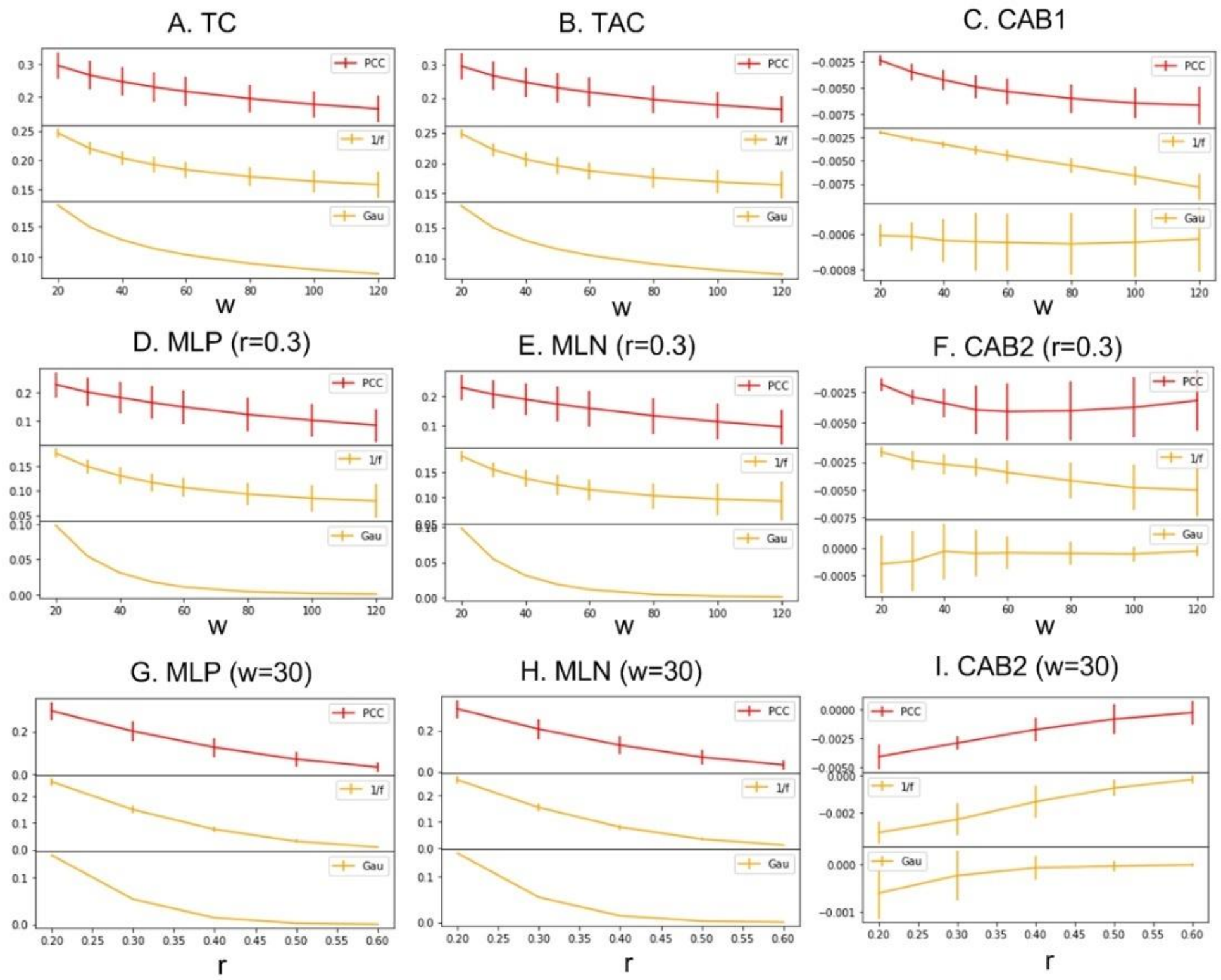

Fig. 4. TCM measures at different embedding vector length (w, the horizontal axis) and r. A, B, C are independent of r. r=0.3 in D, E, and F. w=30 in G, H, and I. Error bars indicate STD of the measures from 20 different samples of each of the three time series.

*Effects of w on TC, TAC, and CAB1*

Fig. 4 shows the evaluation results of TCM with different parameters, w, and r, for the three time series. Both ANOVA and paired-t tests were used to assess the effects of different w values on TC, TAC, and CAB1. ANOVA was used to assess the overall effects of w, while paired-t test was used to assess the TCM measure difference between two different w values. For each time series, TC (Fig. 4A) and TAC (Fig. 4B) both decreased with w (p<3.5e-18, one way ANOVA for each of the three time series). The corresponding changes caused by different w were statistically significant (p<2.4e-5, paired-t test, two tailed for each possible pair of w values). CAB1 of gaussian noise did not show significant changes with w (p=0.97, one-way ANOVA).

*Effects of w on MLP, MLN, and CAB2*

Fig. 4D and 4E show that both MLP and MLN decrease with w. Both ANOVA and paired-t test were used to assess the effects of different w values on MLP, MLN, and CAB2. These assessments were made for each r separately. For each assessed r (from 0.2 to 0.6), MLP showed statistically significant ($p<1.47e-16$, one-way ANOVA, the factor is w) decrease in all three signals. MLN decreases with w ($p<1.45e-15$, one-way ANOVA, the factor is w) for all three signals for all r values. PCC rsfMRI time series showed significant CAB2 changes only for r=0.2 ($p<2.01e-19$). 1/f noise showed significant CAB2 changes (Fig. 4F) due to the change of w ($p<7.94e-14$) when $r<=0.3$. CAB2 of Gaussian noise showed significant changes when r=0.2, 0.5 and 0.6 ($p<1.54e-5$). MLP differences between any two w values for any of the r value were statistically significant ($p<9.85e-5$, two tailed paired t-test). MLN differences between any two w values for any assessed r value were statistically significant ($p<9.86e-5$, two tailed paired t-test). CAB2 of the PCC rsfMRI time series and Gaussian noise did not differ across w for most of r values. CAB2 of 1/f noise showed statistically significant differences in 1/4 of the total number of paired-t test.

*Effects of r on MLP, MLN, and CAB2*

Fig. 4G and 4H show that both MLP and MLN decrease with r. ANOVA and paired-t test were used to assess the effects of r on MLP, MLN, and CAB2 for each w separately. For each assessed w (from 20 to 120), MLP showed statistically significant ($p<4.6e-26$, one-way ANOVA, the factor is r) decrease in all three signals. MLN decreases with r ($p<1.7e-27$, one-way ANOVA, the factor is w) for all three signals. CAB2 differed significantly ($p<2.2e-5$, one-way ANOVA) across r for all three signals except for the random noise when w=80, 100, or 120. MLP was significantly different between two different r values in all three signals for all w values ($p<0.019$, two tailed paired t-test). MLN was significantly different between two different r values in all three signals for all w values ($p<0.017$, two tailed paired t-test). CAB2 was significantly different between two different r values in all three signals for all w values ($p<0.045$, two tailed paired t-test).

*Effects of w on the cross-signal TC, TAC, and CAB1 difference*

TC, TAC, and CAB1 significantly differed across the three signals at all assessed w ($p<0.041$ for all possible two-sample t-test on each of the three property measures for any two of the three signals).

*Effects of w and r on the cross-signal MLP, MLN, and CAB2 difference*

MLP, MLN, and CAB2 can significantly differentiate the three signals for all assessed w and r ($p<0.046$ for all possible two-sample t-test on each of the three property measures for any two of the three signals).

*Mean TCMs*

TCMs calculated with different w and r were very similar though the intensity was different. Both ICC and the correlation analyses showed very similar results for different w and r too. Based on the synthetic data and the result similarity of the in vivo data, I only showed results based on w=30 and r=0.3 below.

Fig. 5 shows the mean TCMs of REST1 (the first LR and RL rsfMRI scans) of all 862 subjects. TC, TAC, MLP, and MLN had very similar image contrast with high value in the cortical region and lower value in subcortical area and white matter (note that most part of white matter has been masked out during TCM calculations in order to save computation time). Similar to the results of the synthetic data, MLP and MLN are shorter than 15, i.e., w/2. CAB1 (Fig. 5C) and CAB2 (Fig.

5F) showed very different contrast. CAB1 was smaller than 1 in the entire brain. Image contrast of CAB1 looks opposite to that of TC, TAC, MLP, or MLN. Cortical CAB1/CAB2 were more negative than white matter and subcortical CAB1/CAB2. TCM maps of REST2 were nearly identical to those of REST1 and therefore not shown.

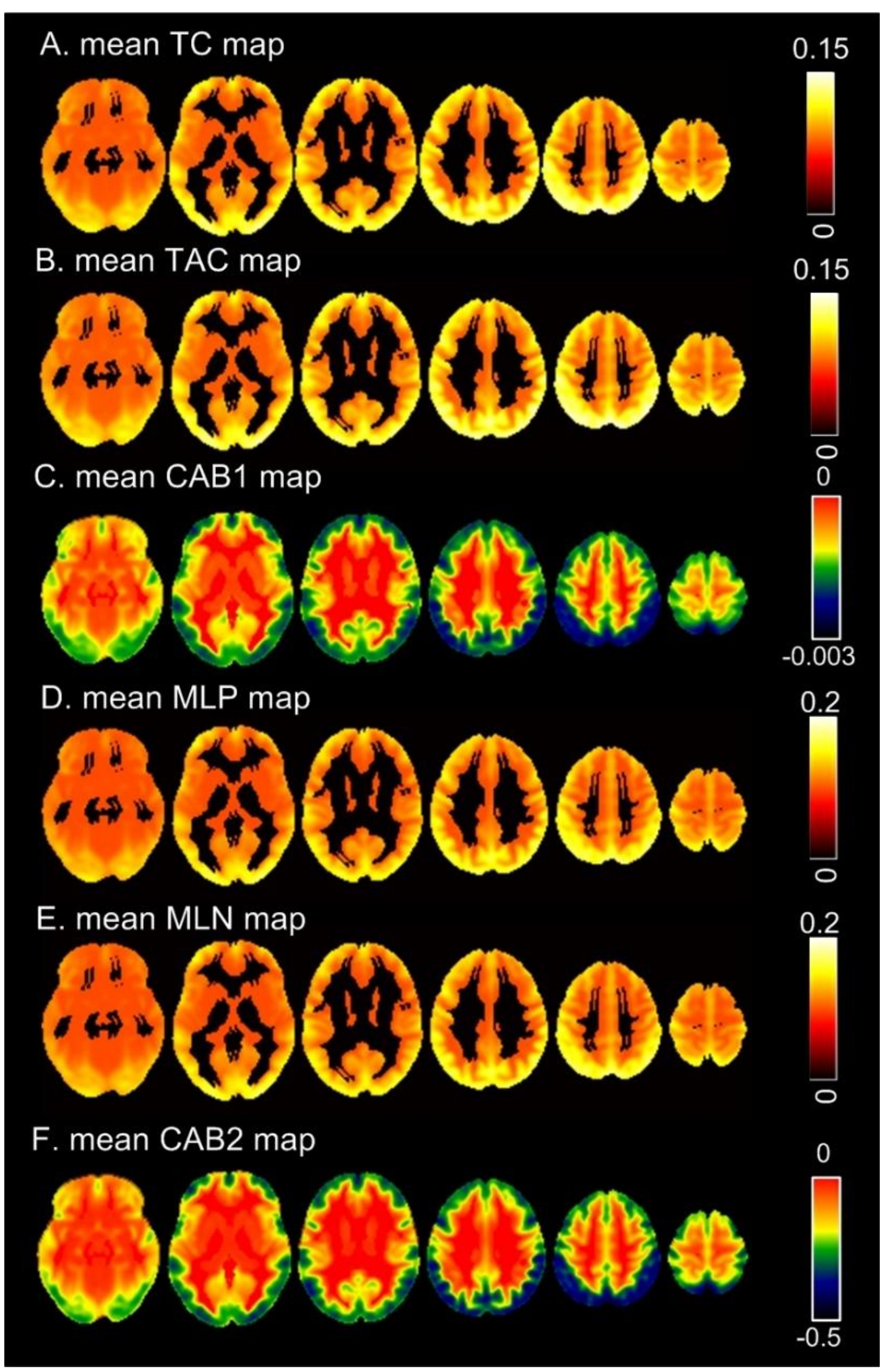

Fig. 5. Mean TCMs of all subjects at the REST1 session. Display window for each map was determined by the colorbar on the right.

*Test-retest analysis results*

Fig. 6 shows the ICC maps thresholded at ICC>=0.3. TC, TAC, MLP, and MLN showed high ICC in the entire brain (note that most part of white matter was masked out during TCM calculation). CAB1 and CAB2 were much less reliable. CAB1 showed moderate ICC values in cingulate cortex, insula, primary visual cortex, orbitofrontal cortex, dorsolateral prefrontal cortex, and inferior temporal lobe. CAB2 was non-stable in nearly the entire brain except for a small part of motor cortex and visual cortex.

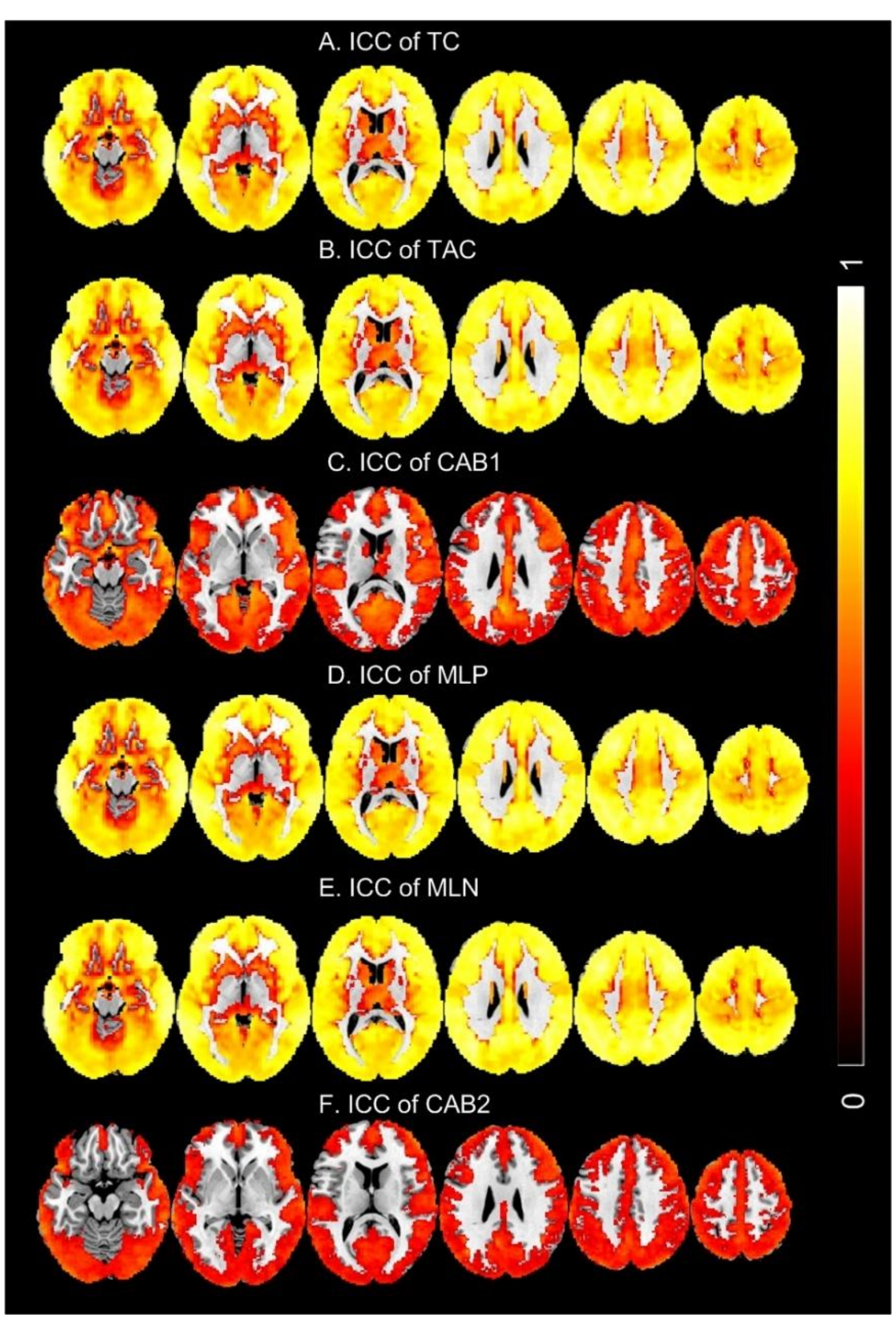

Fig. 6. ICC maps of the six TCMs: A) TC, B) TAC, C) CAB1, D) MLP, E) MLN, F) CAB2. ICC maps were overlayed on top of the MNI standard brain. The cutoff used to threshold the ICC maps was 0.3. The colorbar indicates the display window of the ICC values.

*Biological and cognitive associations of TCMs*

Fig. 7 shows the voxelwise correlations between each of the six TCMs and age. TC/TAC/MLP/MLN showed nearly identical significant ($p<0.05$, multiple comparison corrected using the FWE based method, corresponding to $Z=5.15$) correlations to age in prefrontal cortex, parietal cortex, and temporal lobe. CAB1 and CAB2 (Figs. 7C, 7F) showed nearly identical significant ($p<0.05$, FWE corrected) positive correlations with age in temporal cortex, prefrontal cortex, and lateral parietal cortex. CAB1 presented slightly bigger suprathreshold clusters. The suprathreshold age correlation clusters of CAB1 and CAB2 overlap with but are smaller than those of TC, TAC, MLP and MLN. CAB1 and CAB2 had bigger clusters in inferior prefrontal cortex, temporal pole, and insula but did not show age correlation in posterior cingulate cortex and precuneus.

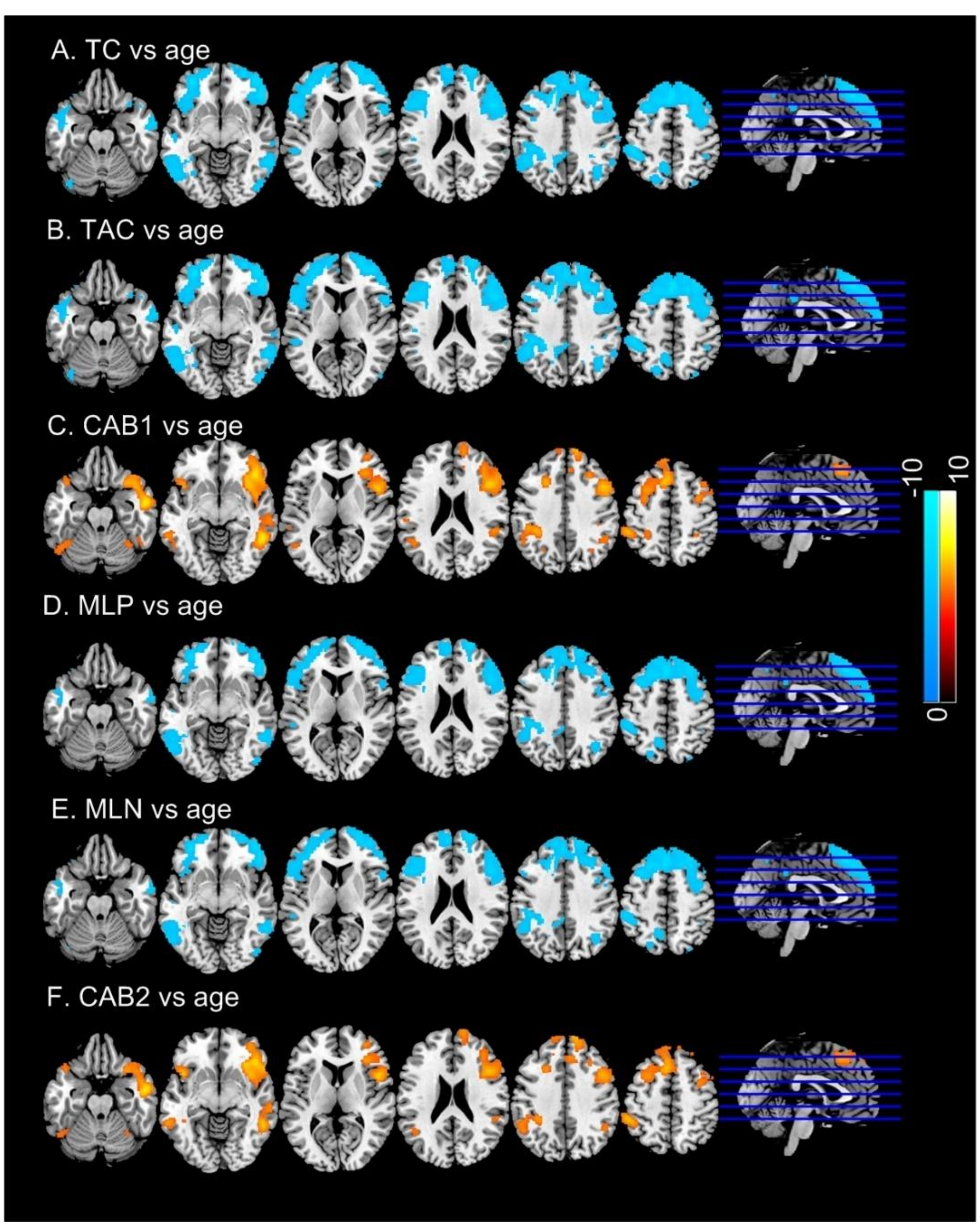

Fig. 7. Correlations of regional TCMs with age: A) TC, B) TAC, C) CAB1, D) MLP, E) MLN, F) CAB2. Significance level was defined by p<0.05 (FWE corrected). Blue color means negative correlation; hot color means positive correlations. Colorbars indicate the display window of the Z-scores of the age vs TCM regressions.

Fig. 8 shows the sex effects on TCMs. Female had lower ($p<0.05$, FWE corrected) TC/TAC/MLP/MLN than male in nearly the entire cortical area, including temporal cortex, insula, parietal cortex, motor cortex, part of prefrontal cortex, and visual cortex. However, female had higher CAB1 (Fig. 8C) and CAB2 (Fig. 8F) in nearly the entire cortical area except for left inferior prefrontal cortex and left superior and middle temporal cortex.

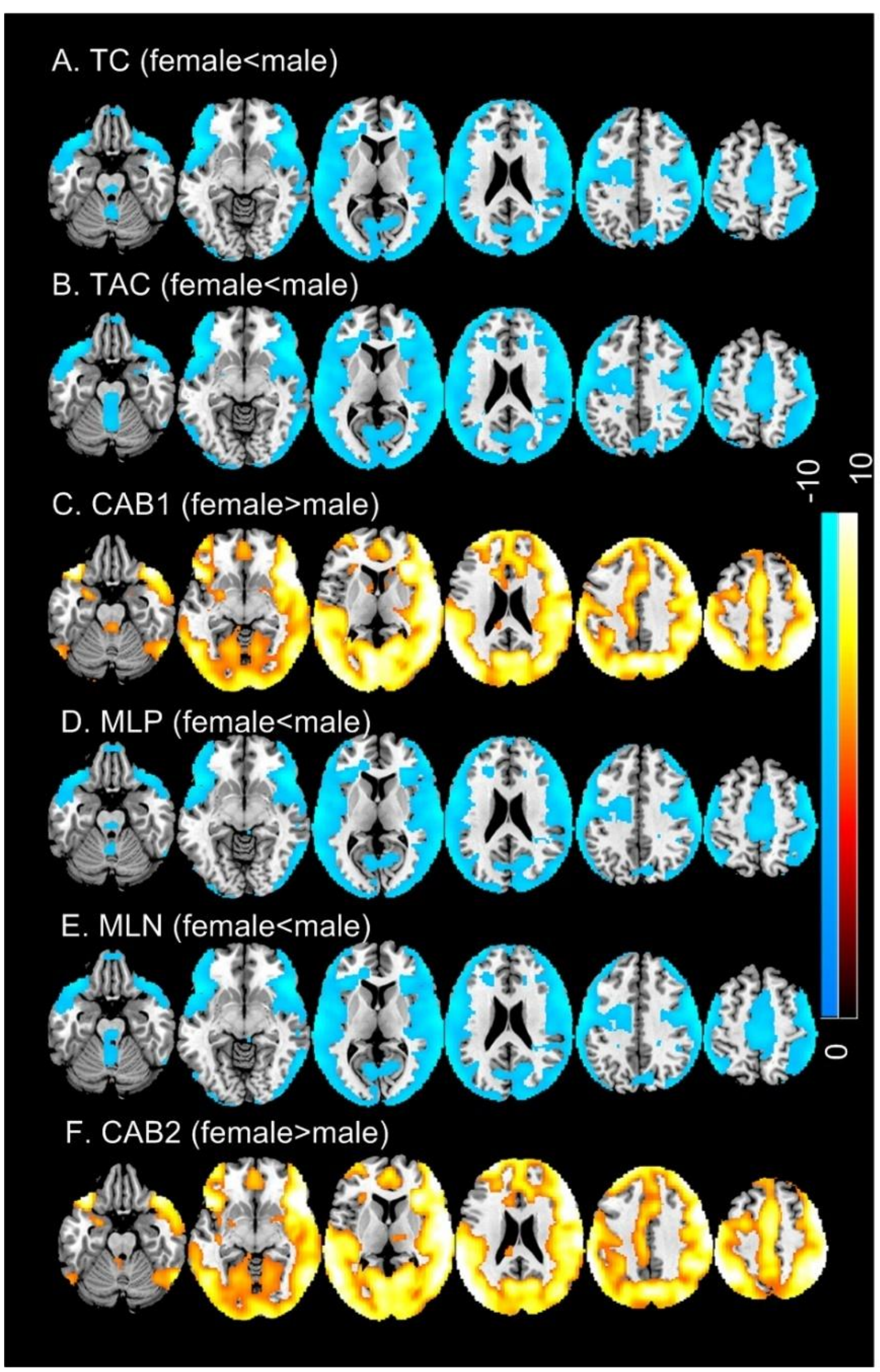

Fig. 8. Sex effects on regional TCMs: A) TC, B) TAC, C) CAB1, D) MLP, E) MLN, F) CAB2. Significance level was defined by $p<0.05$ (FWE corrected). Blue means female lower than male;

hot color means higher in female. Colorbars indicate the display window of the Z-scores of female vs male TCM two-sample t-test.

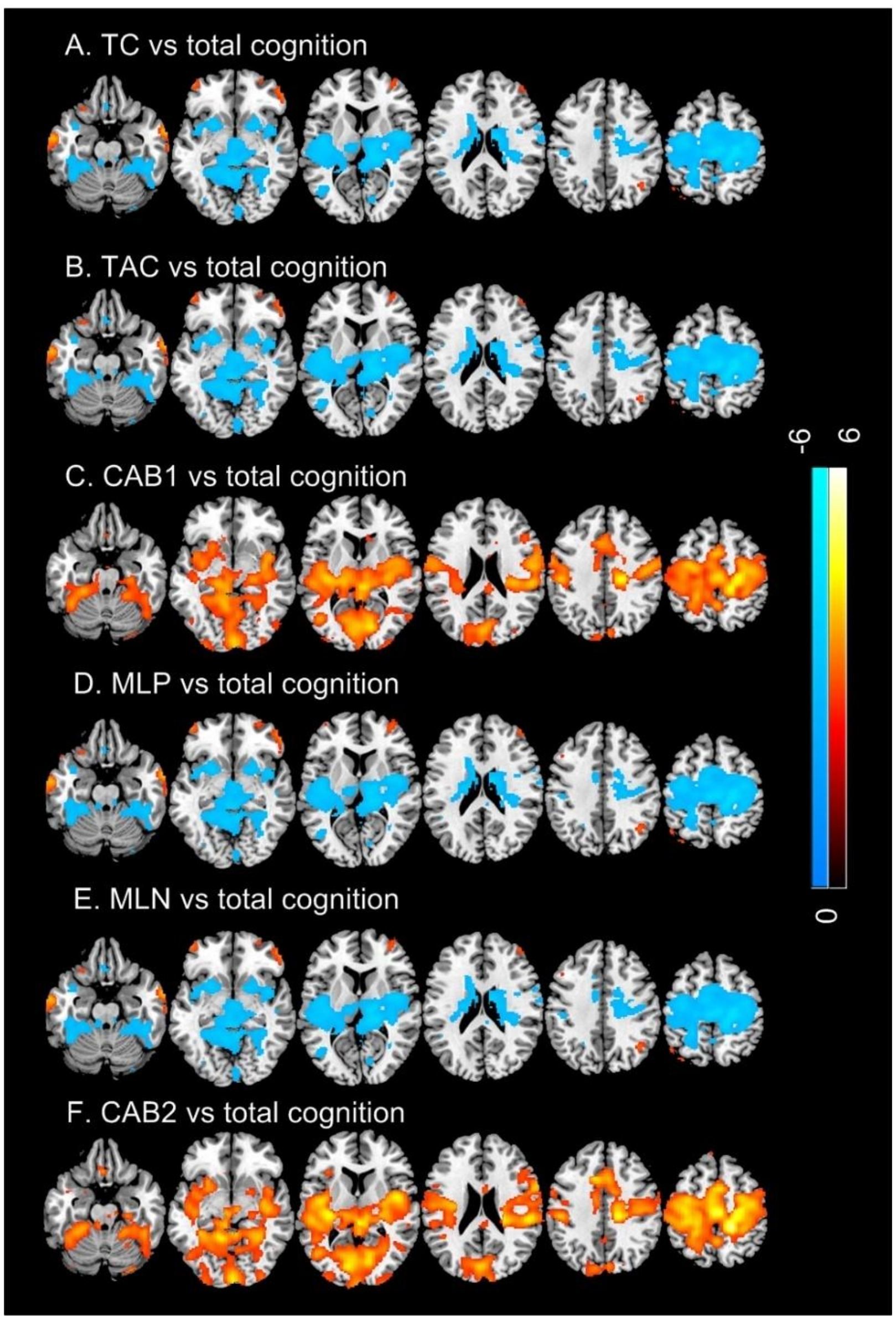

Fig. 9. Cognitive correlations of TCMs: A) TC, B) TAC, C) CAB1, D) MLP, E) MLN, F) CAB2. Significance level was defined by $p<0.05$ (FDR corrected). Blue and hot color indicate negative and positive correlation, respectively. Colorbars indicate the display window of the Z-scores of regression analysis.

Fig. 9 shows the correlations of TCMs with the total cognitive score. TC/TAC/MLP/MLN showed quite similar cognitive correlation patterns (Fig. 9A, 9B, 9D, 9E, $p<0.05$, FDR corrected). The four measures were positively correlated with total cognitive score in temporal cortex, lateral prefrontal cortex, and parietal cortex. They were negatively correlated with total cognitive score in motor cortex, insular, striatum, thalamus, cerebellum, and part of visual cortex. CAB1 and CAB2 (Figs 9C, 9F) only showed positive correlations with the cognitive score in motor cortex, insular, striatum, thalamus, cerebellum, and visual cortex.

## Discussion

I presented a method to exam temporal coherence, temporal incoherence, and their discrepancy in a given time series measured from a dynamic system. The method was dubbed as TCM (temporal coherence mapping) and was based on the correlation coefficient matrix of the temporal embedding vectors. These vectors were extracted to expand the one-dimensional time series to a multiple dimension domain so that the cross-transit state information collapsed in the original one-dimensional space can be better studied. I assessed six typical properties of the temporal coherence matrix. Both synthetic data and in-vivo rsfMRI data were used to evaluate their stability and potential value for neuroscientific research.

Synthetic data were used to evaluate the effects of the length of the embedding window w and the correlation coefficient threshold r on these parameters. TC, TAC, and CAB1 are independent of r. Longer w reduces TC and TAC because longer embedding vectors introduce more variance and subsequently reduces correlation coefficients. CAB1, the difference between TC and TAC was nearly 0 for all assessed signal frequencies and w values. Similar to TC and TAC, MLP and MLN decrease for each assessed signal frequency when w increases. They decrease with frequency when the window length is fixed. Psedo periodic fluctuations appeared when window length is longer than signal cycle. CAB2 was nearly 0 in most of assessed signal cycles and window lengths. Noticeable though still minor to moderate CAB2 values were observed at a few signal frequencies and w values, indicating an artificial MLP/MLN imbalance. This might be caused by the interactions between signal frequency, window length, and the cutoff r. For data with a many frequency components, these minor interactions may be suppressed by the different contributions by the different components. Nevertheless, w and r can be further adjusted to prevent them.

Regarding the comparisons among three different signals, the in-vivo rsfMRI signal had the highest TC, TAC, MLP, and MLN. Different from the uni-frequency sinusoidal, both the rsfMRI and 1/f signal showed a TAC dominance. Gaussian signal did not show statistically significant coherence incoherence imbalance. rsfMRI signal had the biggest TC, TAC, MLP, and MLN and Gaussian signal had the smallest values. TC, TAC, MLP, and MLN decreased with w (embedding vector length) since Pearson correlation coefficient tends to decrease when more data points are included (more data points will increase the chance of having more transit fluctuations which will decrease the correlation coefficient). CAB1 of PCC rsfMRI and 1/f signal were significantly lower than 0, indicating a weak bias toward temporal incoherence. They both decreased with w, suggesting that the potential imbalance can be better observed with big w. CAB2 of rsfMRI decreased with w only when w<=40; CAB2 of the 1/f signal decreased with w. CAB1 and CAB2 of Gaussian signal were not significantly different from 0. MLP and MLN both decreased with r. CAB2 increased (magnitude decreased) with r and became closer to 0. This r dependence was because fewer pairs of embedding vectors could survive the threshold when r increased.

When applied to 865 young healthy subjects' rsfMRI data, I collected the whole brain maps of the six TCM property measures. The maps of TC, TAC, MLP, and MLN had very similar image contrast with high value in the cortical region and lower value in subcortical area and white matter. These findings are consistent with our recent work in brain entropy mapping using rsfMRI from HCP(Wang, 2021b). In grey matter, the highest temporal coherence located in the prefrontal and parietal area which are well known to have predominant slowly fluctuating resting state activity(Buckner and Vincent, 2007; Fransson, 2005; Raichle et al., 2001; Raichle and Snyder, 2007). The high resemblance between TC and TAC maps suggests a tight coupling between coherence and anti-coherence. The high resemblance between MLP and MLN maps suggests that brain regions with coherent transit states staying in parallel have anti-correlated transit states staying for a similar duration of time. The high similarity between TC maps and MLP maps suggests that brain regions that showed higher coherence also had transit states staying parallel for longer time. The high similarity between TAC maps and MLN maps suggests that brain regions that showed higher anti-coherence also had anti-correlated transit states staying anti-correlated for longer time. CAB1 and CAB2 showed very different contrast than the other four TCMs. Lower (more negative) CAB1/CAB2 value was found in the prefrontal and parietal area and close to 0 CAB1/CAB2 was found in inner brain including subcortical region and white matter. Both CAB1 and CAB2 were negative, suggesting that resting brain activity measured by rsfMRI is dominated by anti-coherence. Their values were more negative in grey matter than white matter and subcortical regions, suggesting a larger imbalance of cohernece and anti-coherence in the cortical regions. While this study represents the first effort to characterize coherence and anti-coherence separately in fMRI, the similar distribution patterns of TC and TAC, MLP and MLN, and the patterns of CAB1 and CAB2 suggest a stable imbalance between the macroscopic brain coherence and anti-coherence, which might represent way to coordinate brain excitation and inhibition across large time scale.

All six TCMs were highly reproducible. TC, TAC, MLP, and MLN were reproducible across the brain. CAB1 and CAB2 were reliable mainly in cortical regions and part of cerebellum.

Coherence and anti-coherence maps showed statistically signficant age effects in temporal cortex, prefrontal cortex, precuneus, and parietal cortex. TC, TAC, MLP and MLN all decreased with age. However, CAB1 and CAB2 both increased with age in temporal cortex, insula, prefrontal cortex, and lateral parietal cortex. Because CAB1 and CAB2 were both negative, the postive CAB vs age correlation suggests an age-related decrease of the coherence and anti-coherence imbalance strength. Females showed significantly lower coherence and anti-coherence as measured by TC, TAC, MLP, and MLN. Lower coherent and anti-coherent activity in females was consistent with the higher entropy findings previously reported (Li et al., 2016; Wang, 2021b). Female had less negative CAB1 and CAB2 in the majority of cortex and part of striatum and amygdala, suggesting a weaker imbalance (in terms of imbalance strength) in females compared to males.

LRTCs have been demonstrated to be crucial to high-order brain functions (Botcharova et al., 2014; Buschman and Miller, 2007; Buzsáki and Draguhn, 2004; Dean et al., 2012; Palva et al., 2013; Pesaran et al., 2002; Saleh et al., 2010; Shewcraft et al., 2020; Thut et al., 2012; Womelsdorf et al., 2006a, b; Wong et al., 2016). Our TCM vs total cognitive score correlations are consistent with the literature. Very similar cogitive associations were found in the coherence TCMs and the anti-coherence TCMs, suggesting that coherence and anti-coherence are equally important for brain cognition. The four coherence and incoherence measures were positively correlated with total cognitive score in temporal cortex, lateral prefrontal cortex, parietal cortex, and cerebellum. Negative correlations were found in the sensori-motor system. These regions are well known to have large slowly fluctuating resting state activity (Biswal et al., 1995; Biswal et al., 2010; Damoiseaux et al., 2012; Greicius et al., 2004). Our recent study (Wang, 2021b) has suggested that lower entropy in those regions may indicate a larger capacity of brain function reserve as supported by the negative correlations between entropy and general intelligence and general functionality and the existing speculations that resting state brain activity in these regions may play a role in maintaining and facilitating brain functions (Raichle, 2006; Raichle and Gusnard, 2002; Raichle et al., 2001). Since entropy indirectly reflects coherence, our findings of the positive TCM vs total cognitive score correlations in the frontal, parietal, and cerebellum are consistent with that study (Wang, 2021b). Coherent motor system resting activity has been well characterized using functional connectivity analysis (Biswal et al., 1995). In this study, I found that lower coherence and anti-coherence in sensori-motor networks was associated with better cognitive capability. The associations of TCMs in the sensori-motor system can be explained by the need of fast motor action and sensory information processing or transferring which are often needed during cognitive functions (Leisman et al., 2016). Since less coherent activity indicates a larger number of transit states, the negative sensori-motor system TCM vs cognitive function correlations may also suggest that more transit states are required to support a better cognitive capability. The positive correlation between CAB1, CAB2 and cognition suggests that cognition requires a well balanced coherence and anticoherence in the sensori-motor system, which might be a necessary condition for a quick response of the sensorimotor system.

Pearson correlation coefficient was used to measure the distance between each pair of embedding vectors. This choice was to avoid the effects of single scale which is arbitrary in fMRI. One potential issue is that single scale may change over time because of the potential baseline single drift. This drift is often in low frequency and can be well controlled through temporal filtering or regression, which is a standard step in fMRI data processing. In case of no temporal filtering, the Spearman correlation coefficient can be used as an alternative though it will cost additional computation time for ranking the data.

In an early version, the coherence and anticoherence balance was measured in ratios: TC/TAC and MLP/MLN(Wang, 2021c). I switched to the difference in this version because the ratio may amplify noise due to the division.

The transit temporal correlation (or distance) based coherence and anticoherence assessment can be extended into an inter-regional manner as I shown in a pilot study included in a preprint (Wang, 2021a). It can also be extended to investigate the inter-session or inter-subject temporal coherence and anticoherence.

## Conclusion

Temporal embedding based TCM provides a potentially useful tool to assess brain coherence and anti-coherence and their balance. The coherence/anticoherence measures are stable across repeated measurement. They are correlated with biological and psychological measures. Coherence, anticoherence, and their difference decrease with age in temporal, prefrontal, and parietal cortex. Females had weaker but more balanced coortical coherence and anticoherence. Higher coherence/anti-coherence in temporal cortex, prefrontal cortex, and parietal cortex are correlated with better cognition. Lower coherence in the sensorimotor networks is correlated with better coginition. More balanced sensorimotor coherence and anti-coherence is related to better cognition. These results suggest TCM as a potentially valuable tool for cognitive or translational research.